\documentclass[journal]{IEEEtran}

\usepackage{amsmath,amsfonts}
\usepackage{array}
\usepackage[caption=false,font=footnotesize]{subfig}
\usepackage{textcomp}
\usepackage{stfloats}
\usepackage{url}
\usepackage{verbatim}
\usepackage{graphicx}
\usepackage{xspace}
\usepackage{svg}
\usepackage{booktabs}
\usepackage{tikz}
\usepackage{algorithm}
\usepackage{algorithmicx}
\usepackage{algpseudocode}
\usepackage{listings}
\usepackage{xcolor}
\usepackage{tabularx}
\usepackage{cite}
\def\BibTeX{{\rm B\kern-.05em{\sc i\kern-.025em b}\kern-.08em
    T\kern-.1667em\lower.7ex\hbox{E}\kern-.125emX}}
\usepackage{balance}
\usepackage{enumitem}
\usepackage{siunitx}
\usepackage{fancyvrb}
\usepackage{float}
\floatstyle{ruled}
\newfloat{listing}{tbp}{lol}
\floatname{listing}{Listing}
\newcommand{\codekeyword}[1]{\textcolor[rgb]{0,0.5,0}{\textbf{#1}}}
\newcommand{\codecomment}[1]{\textcolor[rgb]{0.24,0.48,0.48}{\textit{#1}}}
\newcommand{\codeoperator}[1]{\textcolor[rgb]{0.4,0.4,0.4}{#1}}
\newcommand{\codenumber}[1]{\textcolor[rgb]{0.4,0.4,0.4}{#1}}
\newcommand{\codemnemonic}[1]{\textcolor[rgb]{0,0,1}{#1}}
\newcommand{\codeoperand}[1]{\textcolor[rgb]{0.53,0,0}{#1}}
\usepackage[normalem]{ulem}
\usepackage{cleveref}
\usepackage{multirow}

\DeclareSIPrefix\kilo{K}{3}
\setlist[itemize]{leftmargin=10pt,itemindent=0pt,topsep=2pt,partopsep=2.5pt,parsep=1pt,itemsep=1.5pt,listparindent=\parindent{}}

\crefname{section}{Section}{Sections}
\crefname{equation}{Equation}{Equations}
\crefname{figure}{Fig.}{Figures}
\crefname{table}{Table}{Tables}
\crefname{lstlisting}{Listing}{Listings}
\crefname{appendix}{Appendix}{Appendices}
\begin{document}

\title{Improving Indirect Branch Prediction in Interpreters\\via Hardware/Software Co-Design}
\author{Linfeng Zheng, Hiroshi Sasaki}

\newcommand{\name}{\xspace}
\newcommand{\rev}[1]{\textcolor{blue}{#1}}
\newcommand{\look}[1]{\textcolor{red}{#1}}

\maketitle

\begin{abstract}
Interpreters have a large indirect-branch footprint, requiring large predictor capacity for accurate prediction.
We propose a hardware/software co-design in which a hardware lookahead engine, running ahead of the pipeline with software-provided bytecode metadata, supplies interpreter dispatch targets to the frontend.
The engine requires only \qty{1.3}{\kilo\byte} of on-chip storage and changes to $\sim$50 lines of CPython code.
On 15 CPython server workloads, a \qty{14}{\kilo\byte} ITTAGE augmented with the engine reduces bytecode jump MPKI by \num{73.7}\% relative to a \qty{16}{\kilo\byte} ITTAGE baseline, yielding a \num{3.2}\% harmonic-mean IPC speedup.
\end{abstract}

\begin{IEEEkeywords}
Hardware/software interfaces, interpreters, micro-architecture implementation considerations.
\end{IEEEkeywords}

\section{Introduction}
\label{sec:intro}

\begin{figure}[!t]
\centering
\includegraphics[width=\linewidth]{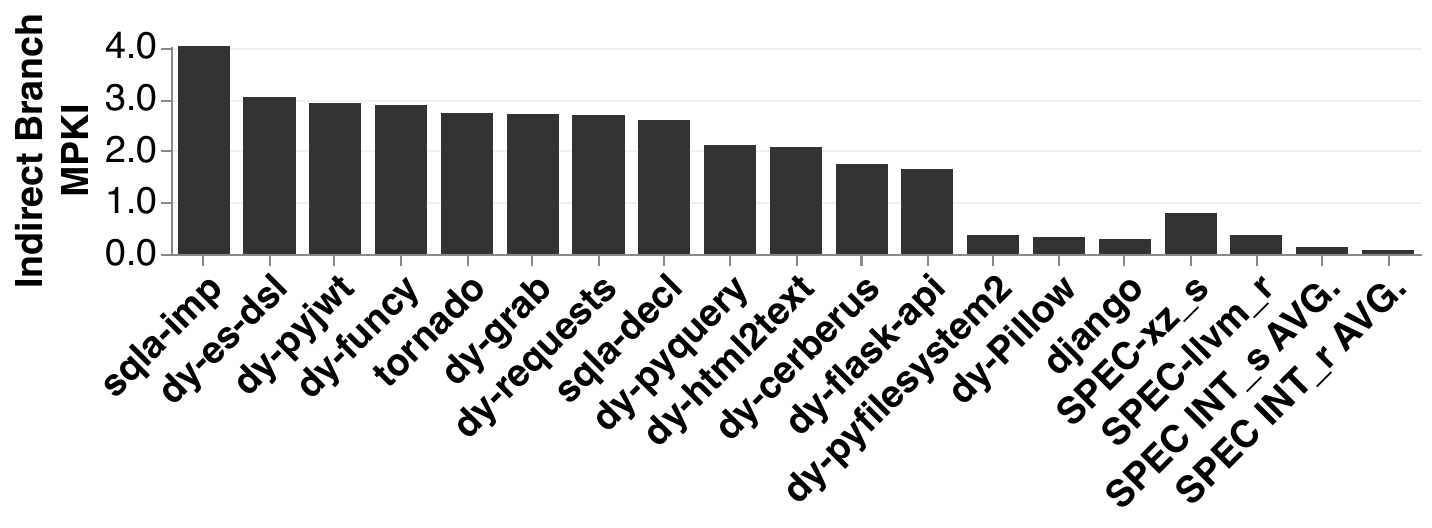}
\caption{\textbf{Commercial-core indirect-branch MPKI.} Measured on an Intel Core i7-13700H P-core for CPython 3.13.14 server and SPEC CPU 2026 integer workloads. The highest-MPKI workloads (\texttt{xz\_s} and \texttt{llvm\_r}) and the average MPKI for SPEC CPU 2026 intspeed and intrate suites are presented.}
\label{fig:commercial_core_mpki}
\vspace{6pt}
\includegraphics[width=\linewidth]{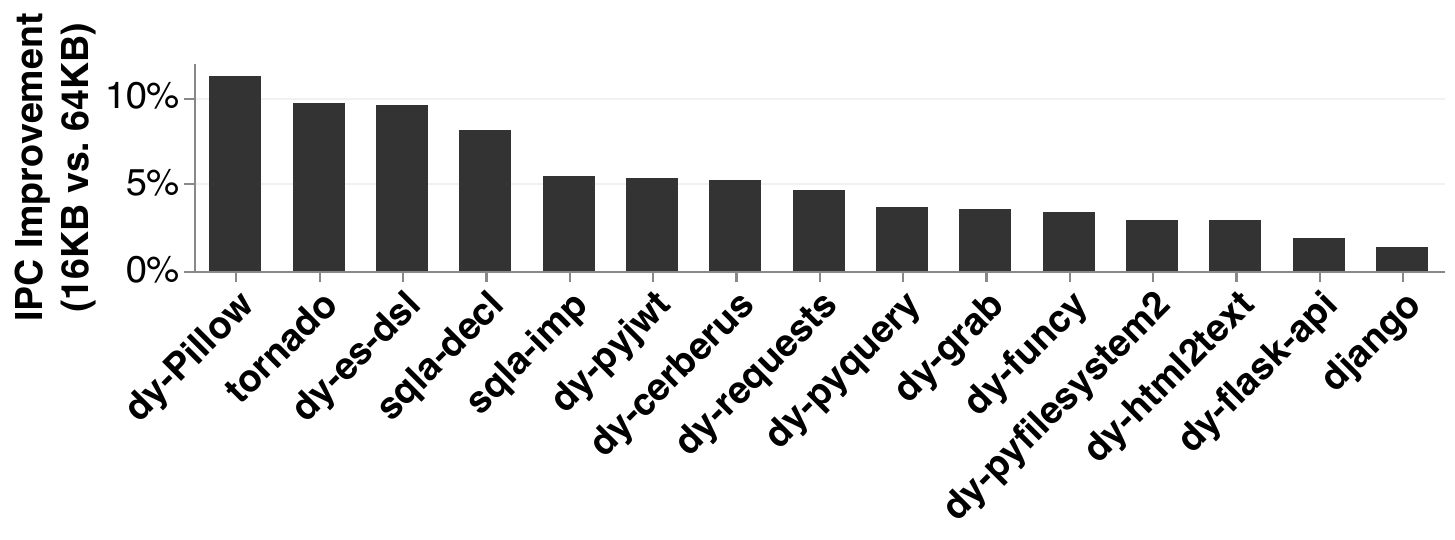}
\caption{\textbf{ITTAGE capacity scaling.} IPC improvement when scaling the baseline indirect branch predictor from \qty{16}{\kilo\byte} to \qty{64}{\kilo\byte}.
The harmonic-mean IPC improvement is 5.1\% (up to 11.2\%), indicating that the workloads' footprint exceeds the \qty{16}{\kilo\byte} capacity.}
\label{fig:ittage_ipc}
\end{figure}

\IEEEPARstart{H}{igh}-level interpreted languages like Python are foundational to modern computing environments.
A well-known performance bottleneck of these languages is frequent indirect branch misprediction.
On a modern Intel Raptor Cove processor with a \num{3072}-entry ITTAGE-style indirect predictor~\cite{seznec201164,li2024indirector}, CPython server workloads exhibit \num{15.2}$\times$ and \num{25.5}$\times$ the indirect branch MPKI of the SPEC CPU 2026 intspeed and intrate averages, respectively (\cref{fig:commercial_core_mpki}).
The experimental CPython~3.13 JIT does not remove the problem: enabling it cuts indirect branches by 10.3\% but the remaining ones mispredict 8.9\% more often, so indirect MPKI falls by only 1.5\%; we therefore evaluate the default, JIT-off configuration.
In gem5, scaling a \num{3392}-entry \qty{16}{\kilo\byte} ITTAGE, comparable to the commercial predictor, to \qty{64}{\kilo\byte}\footnote{\qty{16}{\kilo\byte}: 10 tables (\num{64} entries--\num{1024} entries), history length (\num{16}--\num{1536}); \qty{64}{\kilo\byte}: 15 tables (\num{256} entries--\num{4096} entries), history length (\num{10}--\num{3881}).
The \qty{64}{\kilo\byte} configuration comes from the gem5 repository; the \qty{16}{\kilo\byte} configuration is our own, with table counts, sizes, and history lengths guided by published TAGE/ITTAGE designs.
Scaling entries alone matches the full \qty{64}{\kilo\byte} configuration to within 0.02\% IPC, indicating that capacity, not history length, is the bottleneck.} for the same CPython workloads yields a harmonic-mean IPC improvement of \num{5.1}\% (up to \num{11.2}\%), as shown in \cref{fig:ittage_ipc}.
Dispatch branches account for \num{80.9}\% of the baseline's indirect mispredictions (\cref{fig:results}\subref{fig:res_mpki}).

To tackle this problem, we propose a hardware/software co-design targeting the dispatch indirect branch of interpreters.
The key insight is that bytecodes are laid out as a contiguous array in memory, so the next bytecode for non-control-transfer bytecodes is the next array entry and its dispatch target is determined statically.
With language-specific hints from software (e.g., bytecode format), the hardware can read the next bytecode ahead of time and forward the corresponding handler address to the frontend, reducing reliance on the indirect branch predictor.
For control-transfer bytecodes, whose target cannot be resolved in advance, we extend the existing conditional and indirect branch predictor structures to accept bytecode-level global history for the prediction.

We implement the full system---a $\sim$50-line modification to CPython 3.13.14 and a $\sim$\qty{1.3}{\kilo\byte} lookahead engine in gem5---and evaluate it on \texttt{pyperformance}\cite{pyperformance} and \texttt{DyPyBench}\cite{bouzenia2024dypybench} server workloads.
Under an iso-storage comparison, replacing \qty{2}{\kilo\byte} of a \qty{16}{\kilo\byte} ITTAGE baseline with the engine improves harmonic-mean IPC by \num{3.2}\% (up to \num{7.8}\%).

\section{Background}
\label{sec:background}

Bytecode interpreters such as CPython\cite{python313dis} and V8's Ignition~\cite{v8ignition} compile a program into bytecodes, stored as a contiguous array in memory, and execute them one at a time at the position of a virtual program counter (vPC).
A bytecode consists of an opcode and, optionally, operands such as constants, local variable indices, or branch offsets; the encoding is interpreter-specific (CPython 3.13 uses an opcode byte and an operand byte).
Each opcode has a handler, the native code implementing it, whose address is held in a dispatch table indexed by opcode.
After a sequential (non-control-transfer) bytecode, the vPC advances by the bytecode length; a control-transfer bytecode determines the next vPC at run time from its outcome and target operand.

An interpreter executes bytecodes through a repeated dispatch cycle: it fetches the bytecode at the current vPC, decodes its opcode, indexes the dispatch table to obtain the handler's PC, and jumps to that address via an indirect branch to execute the handler.
Modern interpreters typically replicate this dispatch sequence at the end of each handler---known as threaded dispatch\cite{ertl2003structure}---rather than returning to a centralized switch statement.
As a result, each handler ends with an indirect branch whose target depends on the upcoming bytecode.
Because a single program may use many distinct bytecodes, this dispatch indirect branch has a large set of possible targets requiring a large predictor capacity to avoid predictor aliasing, making it a primary source of indirect mispredictions in interpreted workloads.

\section{Proposed Architecture}
\label{sec:architecture}

\subsection{Overview}
\label{sec:leviathan_overview}

The engine reads the next bytecode from the bytecode array (or predicts it for control-transfer cases), resolves its handler PC from software-provided metadata, and queues the PC for the branch prediction unit (BPU).
Software conveys the bytecode format and handler PCs through system registers and an in-memory metadata table at interpreter startup.
The current vPC is conveyed by a new instruction, \texttt{BR.DISP}, which replaces the dispatch indirect jump.

\subsection{Hardware/Software Interface}
\label{sec:hw_sw_interface}

Given a vPC, the lookahead engine first extracts the opcode from the fetched bytecode.
Two opcode-format registers describe the encoding: \texttt{BCL\_OPCODE\_OFFSET} gives the bit position of the opcode field within the word, and \texttt{BCL\_OPCODE\_WIDTH} gives its width.
The extracted opcode indexes a metadata table, organized like the interpreter's dispatch table but holding per-opcode lookahead metadata rather than a raw handler address.
The interpreter writes the table's base address to \texttt{BCL\_META\_BASE} and the shared high-order handler-PC bits to \texttt{BCL\_HANDLER\_PC\_HI}.

Each metadata-table entry (\Cref{tab:metadata_entry}) is 32 bits wide and corresponds to one opcode.
\texttt{Length} gives the bytecode size used to advance the vPC for sequential bytecodes; \texttt{Type} classifies the bytecode as sequential or as one of three control-transfer kinds (conditional, indirect, or unconditional), determining how the engine resolves the successor; and \texttt{HandlerPCLo} stores the low-order bits of the handler PC.
The lookahead engine reconstructs the full handler PC by concatenating \texttt{HandlerPCLo} with the shared high-order bits stored in \texttt{BCL\_HANDLER\_PC\_HI}.\footnote{This interface directly supports interpreters such as CPython~3.13, where the opcode is the low eight bits of each 16-bit code unit.
Because instructions with inline caches are followed in the code array by non-instruction cache units, Length spans the instruction together with its trailing cache units, so that advancing the vPC by \texttt{Length} lands on the next opcode; the longest instruction reaches \qty{20}{\byte}. Interpreters with operand-scale prefixes (e.g., V8 Ignition) map each opcode–scale pair to its own entry.}

The starting vPC comes from \texttt{BR.DISP}, which replaces the indirect jump ending each handler's dispatch sequence.
\Cref{lst:cpython_dispatch} shows the CPython~3.13 sequence it replaces, where the final \texttt{br x2} jumps to the handler address loaded from \texttt{opcode\_targets[opcode]}.
Architecturally, \texttt{BR.DISP Xt, Xvpc} behaves identically: it transfers control to the native target in \texttt{Xt}.
The second operand \texttt{Xvpc} is visible only to the lookahead engine and carries the vPC of the bytecode being dispatched, i.e., the one whose handler address is in \texttt{Xt}.
The engine consumes \texttt{Xvpc} at commit to synchronize its lookahead with non-speculative execution; from there, it advances through the bytecode stream on its own without further hints.

\begin{table}[t]
\centering
\caption{32-bit metadata entry.}
\label{tab:metadata_entry}
\footnotesize
\begin{tabularx}{\columnwidth}{lcl}
\toprule
\textbf{Field} & \textbf{Bits} & \textbf{Description} \\
\midrule
\texttt{Length} & 5 & Bytecode size in bytes. \\[2pt]
\multirow{2}{*}{\texttt{Type}} & \multirow{2}{*}{2}
  & \texttt{00}: sequential; \texttt{01}: conditional; \\
  & & \texttt{10}: indirect; \texttt{11}: unconditional. \\[2pt]
\multirow{2}{*}{\texttt{HandlerPCLo}} & \multirow{2}{*}{22}
  & Low handler-PC bits, concatenated \\
  & & with the \texttt{BCL\_HANDLER\_PC\_HI} register. \\[2pt]
\texttt{Reserved} & 3 & Reserved for extension. \\
\bottomrule
\end{tabularx}
\end{table}

\begin{listing}[t]
\begin{minipage}{0.48\linewidth}
\begin{Verbatim}[commandchars=\\\{\}, frame=lines, framesep=1mm, fontsize=\scriptsize]
opcode \codeoperator{=} next_instr\codeoperator{->}op.code;
\codecomment{/* next_instr already}
   \codecomment{advanced by the handler */}
\codekeyword{goto} \codeoperator{*}opcode_targets[opcode];
\end{Verbatim}
\centerline{\small (a) Original C code}
\end{minipage}
\hfill
\begin{minipage}{0.48\linewidth}
\begin{Verbatim}[commandchars=\\\{\}, frame=lines, framesep=1mm, fontsize=\scriptsize]
\codemnemonic{ldrh}    \codeoperand{w0}, [\codeoperand{x19}]
\codemnemonic{and}     \codeoperand{w1}, \codeoperand{w0}, \codenumber{#0}\codeoperand{xff}
\codemnemonic{ldr}     \codeoperand{x2}, [\codeoperand{x20}, \codeoperand{x1}, \codeoperand{lsl} \codenumber{#3}]
\codemnemonic{br}      \codeoperand{x2}
\end{Verbatim}
\centerline{\small (b) Corresponding ARM assembly}
\end{minipage}
\caption{Baseline CPython~3.13 dispatch sequence; we replace the final \texttt{br} with \texttt{BR.DISP}, which supplies the dispatched bytecode's vPC.}
\label{lst:cpython_dispatch}
\end{listing}

\subsection{Lookahead Engine Microarchitecture}
\label{sec:microarch}

\begin{figure}[t]
\centering
\includegraphics[width=\linewidth]{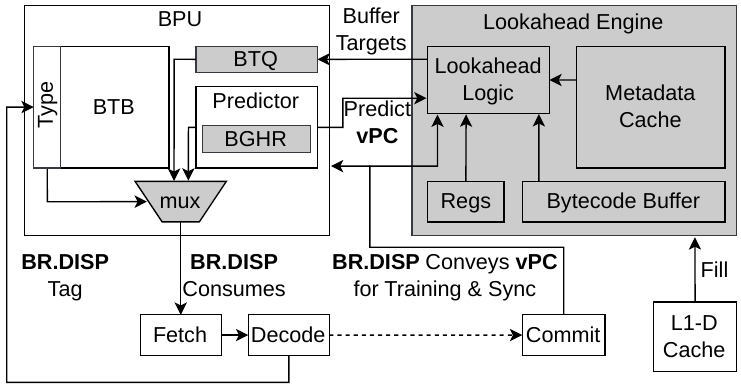}
\caption{\textbf{Lookahead engine microarchitecture.} Decode records \texttt{BR.DISP} in the BTB's branch-type field, fetch reads its target from the BTQ, and commit supplies the vPC that synchronizes lookahead and trains the bytecode-domain predictors. Shaded boxes are added or modified by our design.}
\label{fig:leviathan_overview}
\end{figure}

\Cref{fig:leviathan_overview} shows the lookahead engine and its interface to the frontend.
The engine is a decoupled producer, analogous to the branch predictor that feeds the FTQ in an FDIP frontend, but it walks the bytecode stream rather than the native instruction stream.
Starting from the vPC of the most recent committed \texttt{BR.DISP}, it reads one bytecode, looks up the opcode's metadata to obtain the handler PC and the bytecode \texttt{Type}, and queues the handler PC in the bytecode target queue (BTQ).
It then determines the next vPC: from the bytecode layout for a sequential bytecode, or from the BPU for a control-transfer bytecode, as detailed below. The engine repeats this for successive bytecodes as long as the BTQ has free entries, and the frontend consumes one entry per \texttt{BR.DISP} it fetches.

The engine caches metadata and bytecode locally: a metadata cache holds entries of the in-memory metadata table, and a single-cache-line bytecode buffer holds the line at the current vPC.
Fills come from the L1-D cache through the core's existing port, served only in cycles when the load/store unit issues none, so lookahead never delays the core's own accesses.
Neither buffer requires a dedicated coherence mechanism: the metadata table is written only at interpreter startup, so the metadata cache never goes stale, and the bytecode buffer is kept lazily coherent---a store to the buffered line may cause a misprediction, upon which the engine discards the buffer, refilled from the L1-D cache on the next access.
When the BTQ is full, the engine pauses until a \texttt{BR.DISP} commits and frees a slot.
Capacity is not the limiter: the 256-entry metadata cache holds the entire CPython 3.13 table, and a BTQ beyond 16 entries brings no further gain.

For a sequential bytecode, the next vPC is the current vPC plus \texttt{Length}, so the engine proceeds without prediction.
A control-transfer bytecode needs a predicted successor.
The engine reuses the existing BPU structures---the direction predictor, BTB, and indirect-target predictor---but indexes them with the bytecode vPC instead of a native PC.
It shares the existing ports; an engine lookup stalls frontend lookup for one cycle, but engine lookups are only $\sim$2.7\% of all queries to these structures, so the performance impact is minimal.
The bytecode \texttt{Type} selects the structure: a conditional bytecode queries the direction predictor and BTB, taking \texttt{vPC\,+\,Length} if not-taken and the BTB target if taken; an unconditional bytecode takes the BTB target directly; and an indirect bytecode uses the indirect-target predictor.

Because the engine predicts in the bytecode domain, these lookups use a dedicated bytecode global history register (BGHR), which records recent bytecode control-transfer outcomes, in place of the native GHR.
The engine shifts each predicted outcome into the BGHR speculatively, and a reset recovers by shifting the register back by the number of still-speculative bits, tracked by a 4-bit counter (the shift-back scheme of\cite{skadron2000speculative}); keeping the BGHR separate from the native GHR lets lookahead do this without perturbing native history.

The frontend obtains dispatch targets from the BTQ rather than the native predictor.
When a \texttt{BR.DISP} is first decoded, the engine records it in the branch-type field of its BTB entry; later fetches identify the dispatch branch from that field and read its target from the BTQ, falling back to the native indirect predictor if the entry holds no target.
Each BTQ entry stores the predicted handler PC, the ROB sequence number of its \texttt{BR.DISP} (assigned at rename, so commit can match each retiring \texttt{BR.DISP} to its entry), and a bit marking whether the bytecode is a control transfer.

The engine updates the shared predictors only at commit: each pair of consecutive committed \texttt{BR.DISP} vPCs reveals one resolved bytecode transition, and when the earlier bytecode is a control transfer, the engine trains the corresponding direction, BTB, or indirect-target entry from it and retires the oldest speculative BGHR bit.
Code that executes no \texttt{BR.DISP} never triggers an engine update, so for native applications the predictors behave exactly as in the baseline.

A native-path squash flushes wrong-path instructions but does not prove the bytecode stream wrong: the flushed dispatches, typically re-fetched on the corrected path, want the same targets.
The engine therefore keeps its BTQ contents and only rewinds its in-flight count to exclude the squashed dispatches, so their entries are supplied again when the dispatches re-fetch; because each entry carries its \texttt{BR.DISP} sequence number, each re-fetched dispatch reclaims exactly its own entry, even among in-flight instances of the same static branch.
The engine resets only when a committed \texttt{BR.DISP} resolves to a handler PC that differs from the one the BTQ supplied: this means the bytecode successor was mispredicted, so the engine discards its BTQ contents, speculative BGHR state, and bytecode buffer and restarts from that committed vPC, alongside the pipeline's usual misprediction recovery.

\section{Evaluation}
\label{sec:evaluation}

\subsection{Methodology}

\begin{table}[t]
\centering
\caption{Evaluation setup.}
\label{tab:eval_setup}
\resizebox{\columnwidth}{!}{%
\begin{tabular}{ll}
\toprule
\textbf{Component} & \textbf{Configuration} \\
\midrule
Simulator & gem5 (v25), full-system mode \\[2pt]
Software & CPython 3.13.14 (PGO+LTO), \\
& pyperformance and DyPyBench v2.1.0 \\[2pt]
Core & 3 GHz OoO, 512-entry ROB, 16K-entry BTB \\[2pt]
Pipeline & 6-wide fetch/decode; 8-wide rename/issue/commit \\[2pt]
Frontend & FDIP-style decoupled frontend, 8-entry FTQ \\[2pt]
\multirow{2}{*}{Caches}
  & 64KB L1 I/D (5-cycle), 2MB L2 (12-cycle) \\
  & 8MB shared L3 (34-cycle) \\[2pt]
Conditional BP & 64KB TAGE-SC-L\cite{seznec2014tagescl} \\[2pt]
\multirow{2}{*}{Indirect BP} & \begin{tabular}[t]{@{}l@{}}Baseline: \qty{16}{\kilo\byte} ITTAGE \\
    Proposed: \qty{14}{\kilo\byte} ITTAGE + \qty{1.3}{\kilo\byte} lookahead engine\end{tabular} \\
\bottomrule
\end{tabular}%
}
\end{table}

\begin{table}[t]
\centering
\caption{Per-core on-chip storage added by lookahead engine.}
\label{tab:storage_overhead}
\scriptsize
\begin{tabularx}{\columnwidth}{lXr}
\toprule
\textbf{Component} & \textbf{Configuration} & \textbf{Bytes} \\
\midrule
Metadata cache & 256 entries $\times$ 32-bit metadata & 1024 \\[2pt]
\multirow{2}{*}{BTQ} & 16 entries  & \multirow{2}{*}{148} \\
& \hspace{0.3cm}$\times$ (64-bit PC + 9-bit seq. + 1-bit ctrl.) & \\[2pt]
Bytecode buffer & One 64-byte cache line & 64 \\[2pt]
Bytecode GHR (BGHR) & One 256-bit GHR & 32 \\[2pt]
Configuration registers & 4 registers (2$\times$8b + 64b + 32b) & 14 \\[2pt]
Commit/recovery state & 2 $\times$ 64-bit latches and 4-bit counters & 17 \\
\midrule
\textbf{Total} & & \textbf{1299} \\
\bottomrule
\end{tabularx}
\end{table}

We evaluate the proposed engine using the cycle-level gem5 simulator\cite{Lowe-Power:2020:gem5-20:short} in full-system mode.
We modify CPython~3.13.14 and build it using PGO+LTO.
We evaluate server-oriented workloads from \texttt{pyperformance}\cite{pyperformance}, the standard CPython benchmark suite, and \texttt{DyPyBench}\cite{bouzenia2024dypybench}, a suite of executable real-world Python applications.

\Cref{tab:eval_setup} summarizes the setup: we model a wide AArch64 out-of-order core with an FDIP-style decoupled frontend (an 8-entry FTQ performed best among 8, 16, and 32 for both the baseline and our design; deeper FTQs prefetch further past mispredictions),
and compare our design against a \qty{16}{\kilo\byte} ITTAGE indirect predictor.
Under an iso-storage comparison, we replace \qty{2}{\kilo\byte} of ITTAGE capacity with the $\sim$\qty{1.3}{\kilo\byte} lookahead engine while retaining a \qty{14}{\kilo\byte} ITTAGE fallback.

\Cref{tab:storage_overhead} details the per-core on-chip storage the engine adds.
We assume the BTB's branch-type field has an unused encoding that we assign to \texttt{BR.DISP}, adding no storage.

On hits, the bytecode buffer and metadata cache each take one cycle, so a sequential bytecode yields its handler PC in two cycles and a control-transfer bytecode in three (one more to query the BPU); a miss stalls the engine until the fill completes.
BTQ insertion and consumption take one cycle each, overlapped with subsequent lookups.

\subsection{Evaluation Results}

\begin{figure}
    \centering
    \subfloat[\textbf{IPC improvement.} The proposed engine improves harmonic-mean IPC by 3.2\% (up to 7.8\%) over the \qty{16}{\kilo\byte} ITTAGE baseline.]{\includegraphics[width=\linewidth]{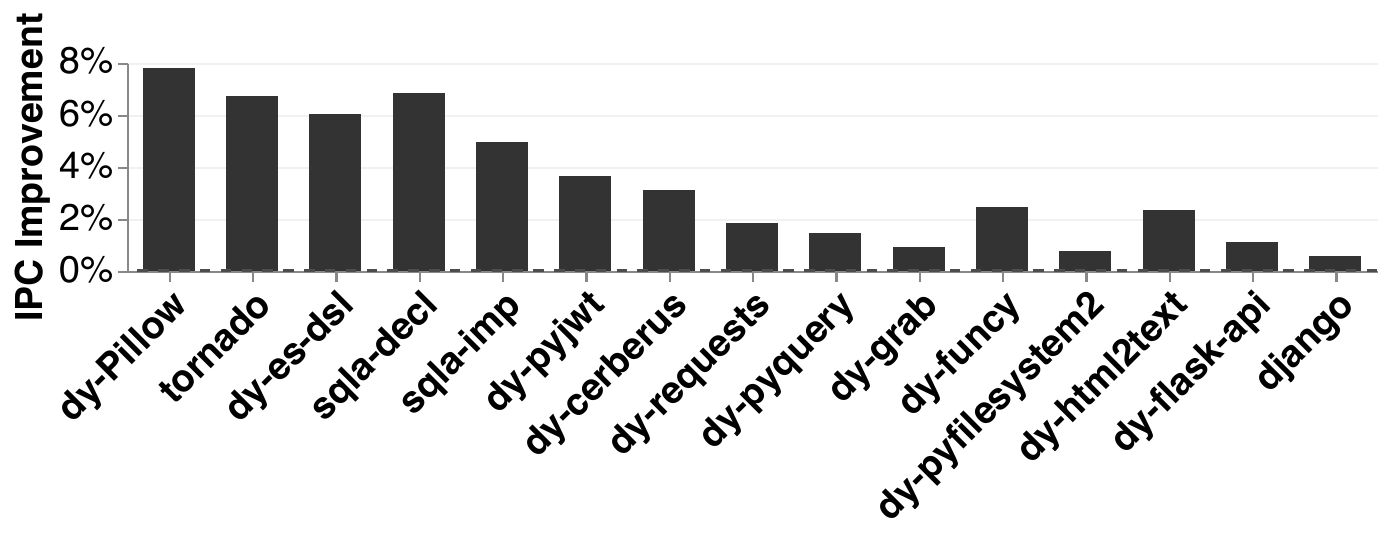}\label{fig:res_ipc}} \\
    \vspace{-2mm}
    \subfloat[\textbf{Branch MPKI breakdown.} Indirect branches are plotted above zero, conditional and unconditional branches below. The proposed engine reduces bytecode jump MPKI by 73.7\% and overall branch MPKI by 23.3\% relative to the baseline.]{\includegraphics[width=\linewidth]{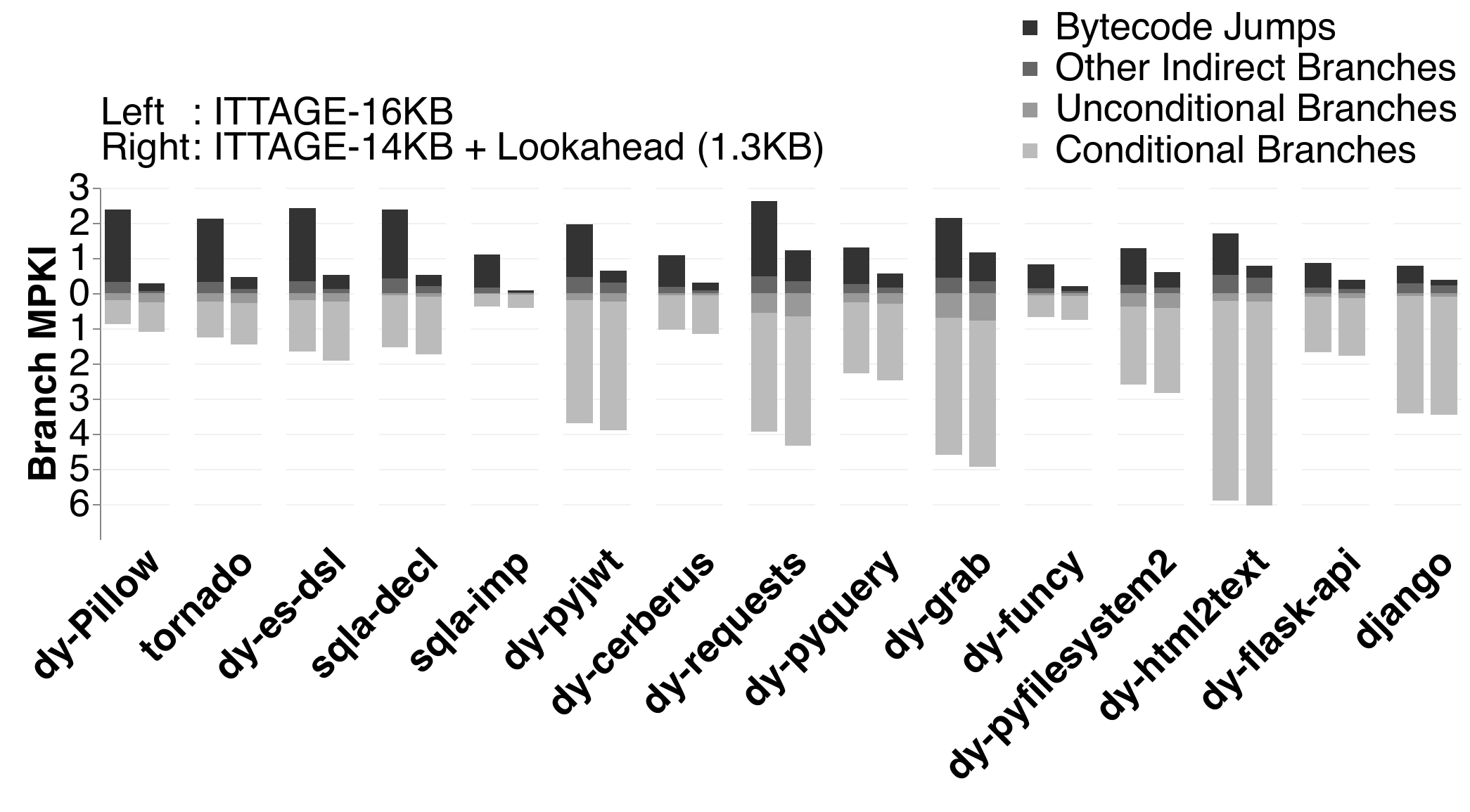}\label{fig:res_mpki}}
    \caption{\textbf{Performance and branch MPKI of the proposed lookahead engine.}}
    \label{fig:results}
\end{figure}

\Cref{fig:results}\subref{fig:res_ipc} reports the IPC improvement over the \qty{16}{\kilo\byte} ITTAGE baseline.
The engine improves harmonic-mean IPC by 3.2\%.
The per-workload gains track the capacity-scaling headroom in \cref{fig:ittage_ipc}: workloads that gain most from enlarging ITTAGE to \qty{64}{\kilo\byte}, such as \texttt{dy-Pillow}, also gain most from the engine (up to 7.8\%), whereas workloads with little scaling headroom, such as \texttt{django}, gain little from either.

The source of the improvement is visible in the MPKI breakdown of \cref{fig:results}\subref{fig:res_mpki}.
The engine reduces bytecode jump MPKI by 73.7\% and overall branch MPKI by 23.3\% relative to the baseline.
The reduction is confined to the bytecode-jump component; the other classes are mostly unchanged (conditional and unconditional branches worsen slightly, as bytecode-domain entries now share their predictor and BTB capacity), as they still use the native predictor.

\section{Related Work}
\label{sec:related_work}

Casey et~al. reduce dispatch mispredictions by replicating handlers and forming superinstructions\cite{casey2007optimizing}, and CPython~3.11+ merges common bytecode pairs and specializes bytecodes\cite{python311whatsnew}; both grow the native footprint, and \cref{fig:commercial_core_mpki} shows that CPython~3.13 still suffers high indirect MPKI.
Rohou et~al. showed that ITTAGE predicts dispatch well until its capacity is exceeded~\cite{rohou2015branch}, the regime we target.
Farooq et~al. index the BTB with a compiler-provided value~\cite{farooq2010vbbi,farooq2012compiler}; for dispatch that value is the opcode, loaded three instructions before the branch (\Cref{lst:cpython_dispatch}), and every (branch, opcode) pair still costs an entry.
Control-Flow Decoupling~\cite{sheikh2012cfd} computes conditional-branch outcomes in a compiler-generated producer loop; a dispatch variant would need no hardware table or predictor sharing, but repeats the load, extract, and table lookup of \Cref{lst:cpython_dispatch} per bytecode and must be re-run whenever specialization rewrites the bytecode array.
Our engine keeps the producer/consumer structure but implements the producer in hardware, adding no instructions and no restructuring.

\section{Conclusions}
\label{sec:conclusion}

We proposed a hardware/software co-design that supplies interpreter dispatch targets through a metadata-guided lookahead engine.
Across 15 CPython server workloads, the design improves harmonic-mean IPC by 3.2\% within the baseline's \qty{16}{\kilo\byte} predictor-storage budget.

\bibliographystyle{ieeetr}
\bibliography{main}

@article{ertl2003structure,
  title={The structure and performance of efficient interpreters},
  author={Ertl, M Anton and Gregg, David},
  journal={Journal of Instruction-Level Parallelism},
  volume={5},
  pages={1--25},
  year={2003}
}

@article{casey2007optimizing,
  title={Optimizing indirect branch prediction accuracy in virtual machine interpreters},
  author={Casey, Kevin and Ertl, M. Anton and Gregg, David},
  journal={ACM Transactions on Programming Languages and Systems},
  volume={29},
  number={6},
  pages={1--36},
  year={2007},
  month={oct},
  publisher={ACM},
  doi={10.1145/1286821.1286828}
}

@inproceedings{farooq2010vbbi,
  title={Value based {BTB} indexing for indirect jump prediction},
  author={Farooq, Muhammad Umar and Chen, Lei and John, Lizy Kurian},
  series={HPCA~'10},
  year={2010}
}

@inproceedings{farooq2012compiler,
  title={Compiler support for value-based indirect branch prediction},
  author={Farooq, Muhammad Umar and Chen, Lei and John, Lizy Kurian},
  booktitle={Compiler Construction},
  series={Lecture Notes in Computer Science},
  volume={7210},
  pages={185--199},
  year={2012},
  publisher={Springer},
  doi={10.1007/978-3-642-28652-0_10}
}

@inproceedings{sheikh2012cfd,
  title={Control-flow decoupling},
  author={Sheikh, Rami and Tuck, James and Rotenberg, Eric},
  series={MICRO~'12},
  year={2012}
}

@inproceedings{seznec201164,
  title={A {64-Kbytes} {ITTAGE} indirect branch predictor},
  author={Seznec, Andr{\'e}},
  booktitle={JWAC-2: Championship Branch Prediction},
  year={2011}
}

@inproceedings{rohou2015branch,
  title={Branch prediction and the performance of interpreters---Don't trust folklore},
  author={Rohou, Erven and Swamy, Bharath Narasimha and Seznec, Andr{\'e}},
  series={CGO~'15},
  year={2015}
}

@article{Lowe-Power:2020:gem5-20:short,
  author       = {Jason Lowe{-}Power and
                  others},
  title        = {The gem5 Simulator: Version 20.0+},
  journal      = {CoRR},
  volume       = {abs/2007.03152},
  year         = {2020},
  url          = {https://arxiv.org/abs/2007.03152},
  eprinttype    = {arXiv},
  eprint       = {2007.03152}
}

@manual{python313dis,
  title        = {{Python 3.13 Documentation: dis --- Disassembler for Python Bytecode}},
  author       = {{Python Software Foundation}},
  year         = {2024},
  url          = {https://docs.python.org/3.13/library/dis.html},
  urldate      = {2026-08-18}
}

@manual{python311whatsnew,
  author       = {{Python Software Foundation}},
  title        = {What's New in Python 3.11},
  year         = {2022},
  note         = {Section ``PEP 659: Specializing Adaptive Interpreter''},
  howpublished = {\url{https://docs.python.org/3.11/whatsnew/3.11.html}}
}

@article{bouzenia2024dypybench,
  title     = {{DyPyBench}: A benchmark of executable python software},
  author    = {Bouzenia, Islem and Krishan, Bajaj Piyush and Pradel, Michael},
  journal   = {Proceedings of the ACM on Software Engineering},
  volume    = {1},
  number    = {FSE},
  pages     = {338--358},
  year      = {2024}
}

@misc{pyperformance,
  author       = {{Python Performance Benchmark Suite contributors}},
  title        = {{The Python Performance Benchmark Suite}},
  howpublished = {\url{https://pyperformance.readthedocs.io/}}
}

@misc{v8ignition,
  author       = {{V8 Project}},
  title        = {Ignition},
  howpublished = {\url{https://v8.dev/docs/ignition}},
  year         = {2016}
}

@inproceedings{li2024indirector,
  title={Indirector: High-Precision Branch Target Injection Attacks Exploiting the Indirect Branch Predictor},
  author={Li, Luyi and Yavarzadeh, Hosein and Tullsen, Dean},
  series={USENIX Security '24},
  year={2024}
}

@inproceedings{seznec2014tagescl,
  title={{TAGE-SC-L} Branch Predictors},
  author={Seznec, Andr{\'e}},
  booktitle={Championship Branch Prediction},
  year={2014}
}

@article{skadron2000speculative,
  title={Speculative updates of local and global branch history: A quantitative analysis},
  author={Skadron, Kevin and Martonosi, Margaret and Clark, Douglas},
  journal={Journal of Instruction-Level Parallelism},
  volume={2},
  pages={1--23},
  year={2000}
}

\end{document}